\documentclass[11pt]{article}
\usepackage{amsfonts}
\usepackage{epsfig}

\newfont{\blackb}{msbm10 scaled\magstep1}

\newfont{\calig}{cmsy10 scaled\magstep1}

\def\text#1{\hbox{#1}}

\newtheorem{theorem}{Theorem}[section]

\newtheorem{remark}{Remark}[section]

\def\be{\begin{equation}}
\def\ee{\end{equation}}
\def\ben{\begin{displaymath}}
\def\een{\end{displaymath}}
\def\baa{\begin{eqnarray}}
\def\eaa{\end{eqnarray}}

\def\ba{\begin{array}}
\def\ea{\end{array}}

\def\g{\gamma}
\def\xb{{\bar{\xi}}}
\def\E{{\cal E}}
\def\Eb{\bar{{\cal E}}}
\def\3{\ss}

\def\ka{\kappa}
\def\l{\lambda}

\def\s{\sigma}
\def\t{\tau}
\def\th{\vartheta}
\def\Th{\Theta}

\def\O{\Omega}

\def\phi{\varphi}

\def\eb{{\bf e}}

\def\Lh{\hat{\L}}

\def\B{{\bf B}}
\def\C{\mathbb{C}}

\def\R{\mathbb{R}}

\def\t0{\Theta_0}
\def\z{{\bf z}}
\def\m{{\bf m}}
\def\la{\label}

\def\c{\cite}
\def\f{\frac}
\def\L{{\cal L}}
\def\p{\partial}
\def\pb{{\bf p}}
\def\qb{{\bf q}}

\def\rb{{\bf r}}
\def\sb{{\bf s}}

\def\tr{{\rm tr}}
\def\0{S}
\def\1{T}

\def\log{\ln}

\begin{document}

\begin{center}{\LARGE Schlesinger system, Einstein equations and hyperelliptic curves}\\
\vskip1.0cm
\hskip7.0cm {\it Dedicated to the memory of Mosh\'e Flato}
\vskip1.0cm
{\large D.~A.~Korotkin}\footnote{E-mail: korotkin@aei-potsdam.mpg.de}\\
\vskip0.5cm
Max-Planck-Institut f\"ur Gravitationsphysik, \\
Schlaatzweg 1, D-14473 Potsdam, Germany\\
\vskip0.5cm
and\\
\vskip0.5cm
{\large V.~B.~Matveev} \footnote{E-mail: matveev@u-bourgogne.fr}\\
\vskip0.5cm
Physique Math\'{e}matique, UFR Sciences et Techniques,\\
Universit\'{e} de Bourgogne, 9 Avenue Alain Savary
B.P. 400, 21011 Dijon Cedex, France
\end{center}

{\bf Abstract.}
We review recent developments in the method of algebro-geometric integration of 
integrable systems related to deformations of algebraic curves. In particular, we discuss 
the theta-functional solutions of Schlesinger system,
Ernst equation, and self-dual $SU(2)$-invariant
Einstein equations.

\vskip1.0cm

\section{Introduction}
\setcounter{equation}{0}

Deep interaction between algebraic geometry of the compact Riemann surfaces and the theory
of integrable systems represent nowadays a well established paradigm of  modern
mathematical physics. The analysis on the compact Riemann  surfaces was   essentially completed in
19th century in the famous classical works by Riemann, Abel, Jacobi, Weierstrass
etc. The penetration of the tools provided by their methods in the theory of integrable systems
started at the same time. Lagrange first solved the equations of motion of the Euler top
by means of the use of the Jacobi elliptic functions.  Then  Carl Neumann 
solved in terms of theta-functions the equations of    geodesic motion
on  ellipsoid  \cite{Neum}. Slightly later Dobriner published
the explicit solution of sine-Gordon equation  expressed in terms of      two-dimensional
theta-functions \cite{Dobr}.
The essential breakthrough was achieved
in 1889 by
Kowalevski \cite{Kova889} who found the new integrable case of the motion of rigid body
solvable in terms of genus 2 algebraic functions. 

The same time it started the development of the spectral theory of the Sturm-Liouville
operators with periodic coefficients.  First important steps in the field were done by
Floquet and  Lyapunov.
In 1919 french matematician Jule Drach wrote the remarkable article  (absolutely forgotten
for next 60 years) devoted to the construction of explicitly solvable Sturm-Liouville
equations associated to hyperelliptic curves of any genus \cite{Drac19}.

One of the subsequent developments of the discovery of the inverse scattering transform (IST) method by
Gardner, Green, Kruskal and Miura \cite{GGKM66} was the complete understanding of the deep
interplay between the classical works listed above. This connection now   became a part of what is
called algebrogeometric approach to the solution of nonlinear differential equations.
This field of activity was initiated in 1974-1976 by the works of
Novikov, Lax, M.Kac, Its, Matveev, Dubrovin, McKean, van Moerbeke and Krichever.

Speaking about the aspects of the theory connected to the explicit solutions of these
equations, one should mention the formulas for solutions of the Korteveg-deVries equation
with periodic initial data obtained in \cite{ItsMat75}. Soon after the same kind of
formulas were derived for numerous other equations integrable via IST method, including
Non-linear Schr\"{o}dinger \cite{NS}, Sine-Gordon   \cite{SG}, Toda lattice  \cite{TL},
Kadomtzev-Petviashvili \cite{KP} and so on. The generic feature of the algebro-geometric
solutions of the systems listed above is that they are parametrised by the fixed algebraic curves
and the associated dynamics is linear on their Jacobians. These results found many
beautiful and unexpected applications in various branches of modern mathematics and physics,
including differential geometry of surfaces \cite{Bobe93} and algebraic geometry
(Novikov conjecture \cite{Shiota}).

It is also relevant to notice that simultaneously with the soliton theory, the
algebro-geometric methods were actively applied in the framework of twistor theory.
 Probably, the main
achievements in this direction were classifications of the instanton and self-dual
monopole configurations.
More recently twistor  methods were applied by Hitchin to classify self-dual $SU(2)$ invariant Einstein manifolds
\cite{Hitc94}. It turned out that the description of most non-trivial subclass of these 
manifolds gives rise to some (solvable in terms of elliptic functions) partial case of 
four-pole Schlesinger system (or, equivalently, Painlev\'e 6 equation).

Technically and conceptually new development was the application of algebrogeometric ideology
to the Einstein equations of general relativity in presence of two commuting Killing vectors
(Ernst equation). The embedding of this equation in the framework of the IST
approach was initiated by Belinskii-Zakharov \cite{BelZak78} and Maison \cite{Mais78}
\footnote{Coming back to the works of classics of differential geometry of 19th century
it is curious to mention that essentially the same Lax pair appeared in the
work of Bianchi \cite{Bian??} in the study of so-called Bianchi congruences.}.
The characteristic feature of the associated zero-curvature representation is the
non-trivial dependence of the associated connection on the spectral parameter. Namely, the
connection lives on the genus zero algebraic curve depending on space-time variables.
This peculiarity entails the drastic change of the construction and qualitative
properties of algebro-geometric solutions which were first obtained in 1988 \cite{Koro88,KorMat89}.
Dynamics in these solutions is generated by hyperelliptic curves with two coordinate-dependent
branch points. The algebro-geometric solutions of the Ernst equation do not possess any
periodicity properties which were inevitable for all KdV-like cases studied before.
Moreover, we can explicitly incorporate in the construction an arbitrary functional
parameter which never appears in the traditional KdV-like 1+1 integrable systems.
Wide subclass of the obtained solutions turns out to be asymptotically flat \cite{Koro93}.
As a simple degenerate case the algebro-geometric solutions contain the whole
class of multisoliton solutions found by Belinskii and Zakharov \cite{BelZak78}.

Simplest elliptic solutions  were studied in  \cite{Koro93}. Despite many interesting properties,
they contain ring-like naked singularities making it difficult to exploit them in a real
physical context. Realistic physical application of the particular family of
the class of algebro-geometric solutions came out from the series of papers of
Meinel, Neugebauer and their collaborators starting from 1993
\cite{MeiNeu94}.
These works were devoted to the investigation of the boundary-value problem corresponding to the
infinitely thin rigid relativistically-rotating dust disc. The embedding of the
dust disc solution in the formulas of \cite{Koro88} as a special solution of genus 2
 was described in \cite{Koro96}.

The recent work \cite{KitKor98} provided the
 unifying framework for results of \cite{Koro88,Hitc94}. In this paper it was shown that an
arbitrary $2\times 2$ inverse monodromy problem with off-diagonal monodromy matrices admits an
explicit solution in terms of multi-dimensional theta-functions.\footnote{Almost simultaneously the same 
inverse monodromy problem was solved in the paper \cite{DIKZ98} motivated by random matrix theory.}
In the subsequent paper \cite{BabKor98} results of \cite{KitKor98} were applied to self-dual $SU(2)$ 
invariant Einstein manifolds, which allowed to considerably simplify the original formulas  of 
Hitchin \cite{Hitc94}.

Existence of close link between the Schlesinger system and Ernst equation, revealed in \cite{KorNic95},
allowed to apply results of \cite{KitKor98} also to Ernst equation \cite{KorMat98} which gave a 
simplification
of original description of \cite{Koro88} and allowed to integrate equations for all metric coefficients
associated to algebro-geometric Ernst potentials.

Our purpose here is to describe results of \cite{KitKor98,BabKor98,KorMat98} in unified framework.

\section{Solutions of Schlesinger system in terms of theta-functions}

Let us briefly remind the origin of the Schlesinger system.
Consider the following linear differential equation for function
$\Psi(\lambda)\in SL(2,\C)$:
\be
\f{d\Psi}{d\g}= \sum_{j=1}^N\f{A_j}{\g-\g_j}\Psi\;,
\la{ls}\ee
where matrices $A_j\in sl(2,\C)$ are independent of $\g$.
Let us impose the initial condition
$\Psi(\g=\infty)=I $.
Function $\Psi(\g)$ defined in this way lives on the
universal covering $X$ of $\C P^1\setminus\{\g_1,\dots,\g_N\}$. The
asymptotical expansion of $\Psi(\g)$ near singularities $\g_j$ is
given by
\be
\Psi(\g)= ( Q_j+ O(\g-\g_j)) (\g-\g_j)^{T_j} C_j\;,
\la{asymp}\ee
where  $T_j$ is traceless diagonal matrix; $Q_j,\;C_j\;\in SL(2,\C)$.
Matrices
\be
M_j=C_j^{-1} e^{2\pi i T_j} C_j, \hskip1.0cm j=1,\dots, N
\la{Mj}\ee
are called the monodromy matrices of system (\ref{ls}).

The assumption of independence of all matrices $M_j$ of the positions of the poles $\g_j$:
$ \p M_j/\p\g_k=0$
is called the isomonodromy condition; it implies the following
dependence of $\Psi(\g)$ on $\g_j$:
\be
\f{\p\Psi}{\p\g_j}=-\f{A_j}{\g-\g_j}\Psi\;.
\la{ls1}\ee
The compatibility condition of (\ref{ls}) and (\ref{ls1}) is equivalent to the
Schlesinger system \c{Schl12} for the residues $A_j$:
\be
\frac{\partial A_j}{\partial\g_i}=
\frac{[A_i,A_j]}{\g_i-\g_j},\;\;\;i\neq j\;,\;\;\;\;\;
\frac{\partial A_i}{\partial\g_i}=
-\sum_{j\neq i}
\frac{[A_i,A_j]}{\g_i-\g_j}.
\label{sch}
\ee
The tau-function   $\tau (\{\g_j\})$ of the Schlesinger system is defined by equations
\be
\f{\p}{\p\g_j}{\rm log}\{\tau\} = \sum_{k\neq j}\f{\tr A_j A_k}{\g_j-\g_k}\;.
\la{tauHj}
\ee
 Compatibility of
equations (\ref{tauHj}) follows from the Schlesinger system.

Certain class of functions $\Psi$, satisfying isomonodromy conditions, was constructed in the paper
\cite{KitKor98}. To describe this construction let us take $N=2g+2$ and  introduce the
hyperelliptic curve $\L$ of genus $\g$ by the equation
\be
w^2=\prod_{j=1}^{2g+2}(\g-\g_j)
\la{L}\ee
with branch cuts $[\g_{2j+1},\g_{2j+2}]$. Let us choose the canonical basis of cycles 
$(a_j,b_j),\;j=1,\dots, g$ such that the cycle $a_j$ encircles the branch cut   $[\g_{2j+1},\g_{2j+2}]$.
Cycle $b_j$ starts from one bank of branch cut $[\g_1,\g_2]$, goes to the second sheet through
of branch cut  $[\g_{2j+1},\g_{2j+2}]$, and comes back to another
bank of the  branch cut $[\g_1,\g_2]$.

The dual basis of  holomorphic 1-forms on $\L$
are given by
$\f{\g^{k-1}d\g}{w},\;\;k=1,\dots,g$.

Let us introduce two $g\times g$ matrices of $a$- and $b$-periods
of these 1-forms: 
\be
{\cal A}_{kj}=\oint_{a_j}\f{\g^{k-1}d\g}{w},\;\;\;\;\;\;\;
{\cal B}_{kj}=\oint_{b_j}\f{\g^{k-1}d\g}{w}.
\la{AB}\ee
The holomorphic 1-forms
\be
dU_k=\f{1}{w}\sum_{j=1}^g ({\cal A}^{-1})_{kj} \g^{j-1} d \g
\la{dUk}\ee
satisfy the normalization conditions
$\oint_{a_j} dU_k=\delta_{jk}$.

The matrices ${\cal A}$ and ${\cal B}$ define the symmetric
$g\times g$ matrix of $b$-periods of the curve $\L$:
$\B= {\cal A}^{-1}{\cal B}$. Now we can also define  the $g$-dimensional theta-function $\Th[^\pb_\qb](\z|\B)$, associated to curve
$\L$, with argument $\z\in\C^g$ and arbitrary complex  characteristic $[^\pb_\qb]$ ($\pb\in\C^g$, $\qb\in\C^g$).
The theta-function has standard periodicity properties with respect to the shift of the argument
on any integer linear combination of vectors $\eb_j\equiv (0,\dots,1,\dots,0)^t$ 
($1$ stands in the $j$th place) and $\B\eb_j$.

Let us cut the curve $\L$ along all basic cycles to get the fundamental polygon   $\Lh$.
For any meromorphic 1-form $dW$ on $\L$ we can define the integral $\int_Q^P dW$,
where the integration contour lies inside of $\Lh$ (if $dW$ is meromorphic,
the value of this integral might also depend on the choice of integration
contour inside of $\Lh$).
The vector of Riemann constants corresponding to our choice of the
initial point of this map is given by the formula (see \cite{Fay}) $K_j=\f{j}{2}+\f{1}{2}\sum_{k=1}^g \B_{jk}$.

The characteristic with components $\pb\in\C^g/2\C^g$,
$\qb\in\C^g/2\C^g$ is called half-integer characteristic: the half-integer
 characteristics are in one-to-one correspondence with the half-periods
$\B\pb+\qb$. To any half-integer characteristic we can assign parity which by definition coincides
 with the parity  of the scalar product $4\langle\pb,\qb\rangle$.

The odd characteristics which will be of importance for us in the sequel correspond to any given subset $S=\{\g_{i_1},\dots,\g_{i_{g-1}}\}$ of $g-1$ arbitrary non-coinciding branch points. The odd half-period associated to the subset $S$ is given by
\be
\B\pb^S+\qb^S= \sum_{j=1}^{g-1}\int_{\g_{1}}^{\g_{i_j}} d U -K
\la{odd}
\ee
where $dU=(dU_1,\dots,dU_g)^t$.
Denote by
 $\Omega_\g\subset\C$ the neighbourhood of the infinite point $\g=\infty$,
such that $\Omega_\g$ does not overlap with projections of all basic cycles
on $\g$-plane.
Let the $2\times 2$ matrix-valued function $\Phi(\g)$ be defined in the
domain $\O_\g$ of the first sheet of $\L$  by the following formula,
\be
\Phi(\g\in\O_\g)=\left(\ba{cc}\phi(\g)\;\;\;\;\; \phi(\g^*)\\
                  \psi(\g)\;\;\;\;\; \psi(\g^*)\ea\right),
\la{Phi}\ee
where functions $\phi$ and  $\psi$ are defined in the fundamental
polygon $\Lh$ by the formulas:
\be
\phi(\g)=\Th\left[^\pb_\qb\right]\left(\int_{\g_1}^\g dU + 
\int_{\g_1}^{\g_\phi}dU\Big|\B\right)\Th\left[^{\pb^\0}_{\qb^\0}\right]\left(\int_{\g_\phi}^\g dU
\Big|\B\right),
\la{phi1}
\ee
\be
\psi(\g)=\Th\left[^\pb_\qb\right]\left(\int_{\g_1}^\g dU + 
\int_{\g_1}^{\g_\psi}dU\Big|\B\right)\Th\left[^{\pb^\0}_{\qb^\0}\right]\left(\int_{\g_\psi}^\g dU
\Big|\B\right),
\la{psi1}
\ee
with two arbitrary (possibly $\{\g_j\}$-dependent) points $\g_{\phi}$,
$\g_\psi\in\L$ and arbitrary constant complex characteristic $\left[^\pb_\qb\right]$; 
$*$ is the involution on $\L$ interchanging the sheets.
An odd theta characteristic $\left[^{\pb^\0}_{\qb^\0}\right]$ corresponds to an arbitrary subset $S$ of $g-1$ branch points via  {\rm Eq.~(\ref{odd})}.

Since domain $\O_\g$ does not overlap with projections of all basic cycles
of $\L$ on $\g$-plane, domain $\L_\g^*$ does not overlap with the boundary
of $\Lh$, and functions $\phi(\g^*)$ and $\psi(\g^*)$ in (\ref{Phi}) are
uniquely defined by (\ref{phi1}), (\ref{psi1}) for $\g\in\O_\g$.

Now choose some sheet of the universal covering $X$,
define new function $\Psi(\g)$ in subset $\Omega_\g$ of this sheet
by the formula
\be
\Psi(\g\in\Omega_\g)=\sqrt{\f{\det \Phi(\infty^1)}{\det \Phi(\g)}}\Phi^{-1}(\infty^1)\Phi(\g)
\la{Psi}\ee
and extend on the rest of  $X$ by analytical continuation.

Function  $\Psi(\g)$ (\ref{Psi}) transforms as follows with respect
to the tracing around basic cycles of $\L$ (by $T_{a_j}$ and $T_{b_j}$
we denote corresponding  operators of analytical continuation):
\ben
T_{a_j}[\Psi(\g)]=\Psi(\g) e^{2\pi i p_j\sigma_3}\;; \hskip1.0cm
T_{b_j}[\Psi(\g)]=\Psi(\g)  e^{-2\pi i q_j\sigma_3}  \;
\een
(by $\sigma_j\;,\;\;j=1,2,3$ we denote standard Pauli matrices).

The following statement proved in the paper \cite{KitKor98} claims that function $\Psi$ 
satisfies condition of isomonodromy, and, therefore, provides a class of solutions of Schlesinger system:

\begin{theorem}\la{theoPsi}
Let $\pb,\qb\in \C^g$ be an arbitrary set of $2g$ constants such that characteristic
$\left[^\pb_\qb\right]$ is not half-integer. Then:
\begin{enumerate}
\item
Function $\Psi(Q\in X)$ defined by (\ref{Psi}) is independent of $\g_\phi$
and $\g_\psi$, and satisfies the linear system (\ref{ls}) with
\be
A_j\equiv {\rm res}|_{\g=\g_j} \left\{\Psi_\g\Psi^{-1}\right\},
\la{Aj}\ee
which in turn solve the Schlesinger system (\ref{sch}).
\item
Monodromies (\ref{Mj}) of $\Psi(\g)$ around points $\g_j$ are given by
\be
M_j= \left(\ba{cc} 0 & -m_j \\m_j^{-1} & 0 \ea\right)\;,
\la{Mj1}\ee
where constants $m_j$ may be simply expressed in terms of $\pb$ and $\qb$
(see \cite{KitKor98}).
\item
The $\tau$-function, corresponding to solution  (\ref{Aj}) of the
Schlesinger system, has the following form:
\be
\tau(\{\g_j\})=[\det{\cal A}]^{-\f 12}
\prod\limits_{j<k}(\g_j-\g_k)^{-\frac 18}\Theta\left[^\pb_\qb\right](0|\B)\;.
\la{tau}\ee
\end{enumerate}
\end{theorem}

\section{Algebro-geometric solutions of Ernst equation}
\setcounter{equation}{0}

Here we are going to describe the application of the construction of previous section to
the Einstein equations with two commuting Killing vectors.
The metric on stationary axially symmetric space-time 
with coordinates $(\rho,z,\phi,t)$ may be represented in the following form:
\be
ds^2=f^{-1}[e^{2k}(dz^2+d\rho^2)+\rho^2 d\phi^2]-f (dt+ F d\phi)^2\;,
\la{metric}\ee
where all metric coefficients $f,k,F$ are assumed to depend only on
$\rho$ and $z$. For this kind of metric the essential part of the Einstein
equations is the  Ernst equation
\be
(\E+\Eb)(\E_{zz}+\f{1}{\rho}\E_\rho+\E_{\rho\rho})=2(\E_z^2+\E_\rho^2)
\la{EE}\ee
for complex-valued function $\E(z,\rho)$ - the so-called Ernst potential. 
The metric coefficients may be restored from 
$\E(z,\rho)$ in quadratures according to the following equations:
\be
f=\Re\E\;,
\hskip1.0cm
F_\xi=2\rho\f{(\E-\Eb)_\xi}{(\E+\Eb)^2}\;,
\hskip1.0cm
k_\xi=2i\rho\f{\E_\xi\Eb_\xi}{(\E+\Eb)^2}\;,
\la{coeff}\ee
where $\xi = z + i\rho$. In terms of the $SL(2,\R)/SO(2)$-valued matrix 
\be
G=\f{1}{\E+\Eb}\left(\ba{cc} 2 & i(\E-\Eb)\\
                            i(\E-\Eb) & 2\E\Eb\ea\right)\;.
\la{g}\ee
the Ernst equation may be equivalently rewritten as follows:
\be
(\rho G_\rho G^{-1})_\rho + (\rho G_z G^{-1})_z = 0\;.
\la{EEg}\ee

The relationship between solutions of Schlesinger system and Ernst equation was
revealed in  \cite{KorNic95}:
\begin{theorem}
Let $\{A_j\}$ be some solution of the Schlesinger system (\ref{sch})
and $\Psi(\g)$ be related solution of equation (\ref{ls}) satisfying the
following conditions:
\be
\Psi^t(\f{1}{\g})\Psi(0)^{-1}\Psi(\g) = I\;,
\la{coset}\ee
\be
\Psi(-\bar{\g})= \overline{\Psi(\g)}\;.
\la{reality}\ee
Let in addition  $\g_j=\g(\l_j,\xi,\xb),\;\l_j\in \C$ for all $j$,
with  function $\g(\l,\xi,\xb)$  given by 
\be
\g=\f{2}{\xi-\xb}\left\{\l-\f{\xi+\xb}{2}+\sqrt{(\l-\xi)(\l-\xb)}\right\}.
\la{gamma}\ee
Then function
\be
G(\xi,\xb)\equiv \Psi(\g=0,\xi,\xb)
\la{sol1}\ee
satisfies Ernst equation (\ref{EEg}).
Metric coefficient $e^{2k}$ of the line element (\ref{metric})
coincides with the $\tau$-function of the Schlesinger system up to explicitly computable factor:
follows:
\be
e^{2k}=C \prod_{j=1}^N \left\{\f{\p\g_j}{\p \l_j}\right\}^{\tr A_j^2/2} \tau\;,
\la{taucon}\ee
where $C$ is an arbitrary constant.
\la{ErnstSchl}
\end{theorem}

Exploiting  the relationship between Schlesinger system and Ernst equation
given by the theorem \ref{ErnstSchl}, and making use of theta-functional solutions of the 
Schlesinger system provided by the theorem
\ref{theoPsi}, we can obtain solutions of Ernst
equation in terms of theta-functions.  The necessary additional work to do is to
choose the parameters of the construction (i.e. the constants
$\l_j$ and vectors $\pb,\;\qb$) to provide the constraints
(\ref{coset}) and (\ref{reality}). To get these constraints
fulfilled we have to assume that the
curve $\L$ is invariant under the holomorphic involution
$\sigma$ acting on every sheet of $\L$ as
$\g\rightarrow \g^{-1}$, 
and anti-holomorphic involution $\mu$ acting on every sheet of $\L$ as
$\g\to -\bar{\g}$.
Constraints (\ref{coset}) and (\ref{reality}) turn out to be
compatible with each other only if the genus $g$ is odd:
\be
g = 2g_0 - 1 \;.
\la{gg0}\ee
Let us enumerate the branch points $\g_j,\;j=1,\dots, 4 g_0$ in such an
order that $\g_j=\g_{j+2g_0}^{-1}\;,\; j=1,\dots,2g_0 $
and for some $k\leq g_0$ we have
\ben
\g_{j}\in i\R\;,\hskip0.5cm 1\leq j\leq 2k\;;\hskip0.8cm
\g_{2j+1}+\bar{\g}_{2j+2} =0\;,\;\;\;2k+1\leq j\leq 2 g_0-1 .
\een

Choosing the canonical basis of cycles on $\L$ in appropriate way \cite{KorMat98}, 
we can easily derive the conditions on constants
$\pb$ and $\qb$ imposed by reductions  (\ref{coset}) and (\ref{reality}) (see \cite{KorMat98}). Then relation
(\ref{sol1}) allowes then to express the Ernst potential in terms of theta-functions associated to curve $\L$.
However, much simpler description may be acheived if we make use of
invariance of curve $\L$ under involution $\s$ and
exploit the spectral parameter $\l$ from (\ref{gamma}) instead of $\g$.
Curve $\L$ may
be realised as fourfold covering of $\l$-plane. Four sheets of this covering will be enumerated 
in such a way that the 
involution $\s$ interchanges the sheets $1\leftrightarrow 3$ and
$2\leftrightarrow 4$; involution $*$ interchanges the sheets
$1\leftrightarrow 2$ and $3\leftrightarrow 4$.

Now we consider the new hyperelliptic curve $\L_0$ of genus $g_0$
defined by   equation
\be
\nu^2=(\l-\xi)(\l-\xb)\prod_{j=1}^{2g_0} (\l-\l_j)\;.
\la{L0}\ee
We draw the ``moving'' branch cut between branch points $\xi$ and $\xb$ and $g_0$ immovable branch cuts
 $[\l_{2j-1}, \l_{2j}]$. The canonical basis of cycles on  $\L_0$ is chosen such that 
cycle $a_j$ encircles the branch cut  $[\l_{2j-1}, \l_{2j}]$. Cycle $b_j$ starts on the left bank 
of the cut $[\xi,\xb]$, goes on the second sheet via the branch cut   $[\l_{2j-1}, \l_{2j}]$ and returnes 
back.

Introduce the  dual basis $dV=(dV_1,\dots,dV_{g_0})^t$ of holomorphic
1-forms on $\L_0$  by
\be
dV_j=\f{1}{\nu}\sum_{k=1}^{g_0} ({\cal A}_0^{-1})_{jk}\l^{k-1} d\l\;,\hskip0.5cm
j=1,\dots, g_0\;,
\la{dVj}\ee
where
\be
({\cal A}_0)_{kj}\equiv \oint_{a_j^0} \f{\l^{k-1} d\l}{\nu}\;,
\hskip0.5cm j,k=1,\dots, g_0\;
\la{A0}\ee
The matrix of $b$-periods of $\L_0$ will be denoted by $\B_0$.

Curve $\L$ is twofold non-ramified covering $\Pi$  of $\L_0$,
such that points of $\L$ related by the involution $\sigma$ project onto
the same point of $\L_0$. Anti-involution $\mu$ inherited from $\L$ acts
on every sheet of $\L_0$ as $\l\rightarrow\bar{\l}$.

Reduction (\ref{coset}) allows to
alternatively express function $\Psi$     
 in terms of theta-functions associated to the curve $\L_0$.
Namely, denote by $\O_\l$ the  neighbourhood of the point $\l=\infty$
being the projection of the domain $\O_\g$ into  $\l$-plane.
Define function $\Phi_0(\l\in\O_\l)$ in the subdomain $\O_\l$ of
the first sheet of $\L_0$ by the following formula:
\be
\Phi_0(\l\in\O_\l)=\left(\ba{cc}\phi_0(\l)\;\;\;\;\; \phi_0(\l^*)\\
                  \psi_0(\l)\;\;\;\;\; \psi_0(\l^*)\ea\right),
\la{Phi0}\ee
where involution $*$ inherited on $\L_0$ from $\L$ interchanges the
$\l$-sheets of $\L_0$;
\be
\phi_0(\l) = \Th\left[^\rb_\sb\right]\left(\int_\xi^\l dV \Big|\B_0\right)\; ,
\hskip1.0cm
\psi_0(\l) = \Th\left[^\rb_\sb\right]\left(\int_\xb^\l dV \Big|\B_0\right)\; ,
\la{phipsi0}\ee
\be
\phi_0(\l^*) = -i\Th\left[^\rb_\sb\right]\left( -\int_\xi^\l dV\Big|\B_0\right)\; ,
\hskip1.0cm
\psi_0(\l^*) = i\Th\left[^\rb_\sb\right]\left(-\int_\xb^\l dV\Big|\B_0\right)\; ,
\la{phipsi01}\ee
for $\l\in\O_\l$. Constant vectors $\rb,\;\sb\in \C^{g_0}$ satisfy the following
reality conditions:
\be
\rb\in \R^{g_0}\;,
\la{abj}\ee
\be
\Re\sb_j = \sum_{l=1}^{g_0}\f{\rb_l}{2}\;,\hskip0.5cm  1\leq j\leq k\;;
\hskip1.0cm  \Re\sb_j = \sum_{l=1,\;l\neq j}^{g_0}\f{\rb_l}{2}\;
,\hskip0.5cm k+1 \leq j\leq g_0 \;.
\la{bbj}\ee

which  may be equivalently rewritten as
\be
\Re(\B_0\rb+\sb) = 0\;.
\la{rcalt}\ee

It turns out that if we choose basic cycles on $\L_0$ and $\L$ in consistent way, namely,
such that \cite{KorMat98}:
\ben
\Pi(a_1)=-(a_1^0+\dots+a_{g_0}^0)\;,\hskip0.6cm
\Pi(b_1)=-2b_1^0\;,\hskip0.6cm
\Pi(a_j)=a^0_j\;,\hskip0.6cm
\Pi(b_j)=b^0_j-b_1^0\;,
\een
and, moreover, impose the following relations between constants $\pb,\qb$ and $\rb,\sb$:
$$
p_1=-\sum_{l=1}^{g_0} r_l\;, \hskip0.3cm
q_1=-2 s_1\;,
$$
\be
p_j=-p_{j+g_0-1}= r_j\;,\hskip0.3cm
q_j=-p_{j+g_0-1}= s_j - s_1 \;,  \hskip0.3cm
2\leq j\leq g_0\;,
\la{pqrs}\ee
the function $\Psi$ (\ref{Psi}) in $\O_\l$ may alternatively be written as follows:
\be
\Psi(\l\in \O_\l)=\sqrt{\f{\det \Phi_0(\infty^1)}{\det \Phi_0(\l)}}\Phi_0^{-1}(\infty^1)\Phi_0(\l)\;.
\la{Psi00}
\ee
Again, we extend it to the universal covering $X$ by analytical continuation to
get function $\Psi(P\in X)$.
The proof of coincidence of funtions (\ref{Psi00}) and (\ref{Psi}) may be
obtained by verifying the coincidence of  monodromies of these functions  
around the singularities \cite{KorMat98}.

Now we are in position to formulate the following statement:
\begin{theorem}
\la{main}
Let  $\rb,\sb\in \C^{g_0}$ be arbitrary  constant vectors satisfying
reality conditions (\ref{abj}), (\ref{bbj}).
Then the function
\be
\E(\xi,\xb)=
\frac{\Th\left[^\rb_\sb\right]\left(\int_\xi^{\infty^1}dV \Big|\B_0\right)}
{\Th\left[^\rb_\sb\right]\left(-\int_\xi^{\infty^1}dV|\B_0\right)}
\la{E0}\ee
 solves the  Ernst equation (\ref{EE}).

\end{theorem}

{\it Proof.}
One can represent matrix $\Psi(\l=\infty^1)$,  given by (\ref{Psi00}),
in the form (\ref{g}) with
\ben
\E=-i\f{\phi_0(\infty^1)}{\phi_0(\infty^2)}\;,
\een
which leads to  (\ref{E0}) after substitution of (\ref{phipsi0}).
\vskip0.5cm
\begin{remark}\rm \cite{Koro96,KorMat98}
It turns out that, up to trivial limiting prcedure,
 class of solutions described by simple expression (\ref{E0}) coincides with
the class of algebro-geometric solutions of Ernst equation originally obtained in \cite{Koro88} in more
complicated form. 
\end{remark}

Now it arises the non-trivial problem of integration of equations (\ref{coeff})
for the metric coefficients $F$ and $k$. It turns out that the arising quadratures may be
explicitly resolved; for metric coefficient $F(\xi,\xb)$ we get the following formula \cite{KorMat98}:
\be
F=\f{2}{\Re\E}\Im\left\{\sum_{j=1}^{g_0}({\cal A}_0^{-1})_{g_0 j}\f{\p}{\p z_j}
\log\Th\left[^\rb_\sb\right]\left(-\int_\xi^{\infty^1}dV \Big|\B_0\right)
\right\}
\la{AAA}\ee
where matrix ${\cal A}_0$ is the matrix of $a$-periods of holomorphic 1-forms (\ref{A0}); 
$\f{\p\Th}{\p z_j}$ denotes derivative of theta-function with
respect to its $j$th argument.

More complicated problem of determination of coefficient $k(\xi,\xb)$ can be resolved using
the expression (\ref{tau}) for the $\tau$-function of Schlesinger system and relation (\ref{taucon}) 
between the $\tau$-function and the
coefficient $e^{2k}$ (remind that for algebro-geometric solutions of the Schlesinger system described in the
previous section  $\tr A_j^2=1/8$):
\be
e^{2k}=
\f{\Th\left[^\pb_\qb\right]\left(0 |\B\right)}
{\sqrt{\det {\cal A}_0}}\prod_{j=1}^{2g_0}|\l_j-\xi|^{-1/4}\;,
\la{confac}\ee
where $\B$ is the matrix of $b$-periods of curve $\L$ (\ref{L});
vectors $\pb,\qb\in \C^{2g_0-1}$ are expressed in terms of vectors $\rb,\sb\in\C^{g_0}$ via relations
(\ref{pqrs}).

To get (\ref{confac}) we should also make use of the following simple relation between
determinants of matrices of $a$-periods of holomorphic 1-forms
 of curves $\L$ and $\L_0$:
\be
\det {\cal A}=const\;\rho^{g_0^2}\det {\cal A}_0\;,
\la{AA0}\ee
where the constant is independent of $(\xi,\xb)$.

As usual, degeneration $\l_{2j+1},\;\l_{2j+1}\to\ka_j\in\R$ of curve $\L_0$ into genus zero curve with double points
leads after appropriate choice of constants $\rb,\sb$ to the 
family of Belinskii-Zakharov multisoliton solutions.
In particular, to get Kerr-NUT soultion  one has to take
$g_0=2$ and
\ben
r_j=\f{n_j}{2}\;;\hskip0.5cm s_j=\f{n_j}{4}+i\alpha_j,\hskip0.5cm
\alpha_j\in \R\;, \hskip0.5cm n_j=\pm 1\;, \hskip0.5cm j=1,2\;.
\een
Then (\ref{E0}) turns into \c{Koro88}:
\ben
\E=\f{1-\Gamma}{1+\Gamma}\;, \hskip0.7cm
\Gamma^{-1}=\f{i(a_1-a_2)}{1-a_1 a_2 +i(a_1+a_2)}X+
\f{1+ a_1 a_2}{1-a_1 a_2 +i (a_1+a_2)}Y\;,
\een
where $a_j= n_j e^{-2\pi i \alpha_j}$, $j=1,2$, and
\ben
X=\f{1}{\ka_1-\ka_2}
\Big\{\sqrt{(\ka_1-\xi)(\ka_1-\xb)}+\sqrt{(\ka_2-\xi)(\ka_2-\xb)}\Big\}
\een
\ben
Y=\f{1}{\ka_1-\ka_2}
\Big\{\sqrt{(\ka_1-\xi)(\ka_1-\xb)}-\sqrt{(\ka_2-\xi)(\ka_2-\xb)}\Big\}
\een
are prolate ellipsoidal coordinates.

\section{Self-dual $SU(2)$ invariant Einstein metrics and an example of Frobenius manifold}
\setcounter{equation}{0}

Here we are going to describe two other applications of the simplest non-trivial (elliptic) 
version of construction of section 2. Take $N=4$; as is well-known, the Schlesinger system is in this case
equivalent to Painlev\'e 6 equation. In sect. 2 we got a class of solutions of the four-pole Schlesinger system 
satisfying conditions $\tr A_j^2=1/8$ and $\sum_{j=1}^4 A_j =0$. 
Without loss of generality we can put $\g_4=\infty$.
Then the four-pole Schlesinger 
system may be reformulated in terms of new dependent variables $\Omega_1,\Omega_2,\Omega_3$ defined by
\be
\Omega_1^2=-(\f{1}{8} +\tr A_1 A_2)\;, \hskip0.5cm
\Omega_2^2=\f{1}{8} +\tr A_2 A_3\;, \hskip0.5cm
\Omega_3^2=\f{1}{8} +\tr A_1 A_3
\la{defOm}\ee
and new independent variable - the module
\be
x=\f{\g_3-\g_1}{\g_3-\g_2}\;
\la{x}\ee
of curve $\L$.
If we choose  signes of functions $\Omega_j$ in appropriate way, they satisfy the following system of equations
\cite{Hitc94}:
\be
\f{d\Omega_1}{d x}= -\f{\Omega_2\Omega_3}{x(1-x)}\;, \hskip0.5cm
\f{d\Omega_2}{d x}= -\f{\Omega_3\Omega_1}{x}\;, \hskip0.5cm
\f{d\Omega_3}{d x}= -\f{\Omega_1\Omega_2}{1-x}\;, \hskip0.5cm 
\la{eqOm}\ee
together with condition 
\be
-\Omega_1^2+\Omega_2^2+\Omega_3^2=\f{1}{4}\;.
\la{int}\ee
We can mention at least two different geometrical applications of the system (\ref{eqOm}):
\begin{enumerate}\rm
\item
{\it Self-dual $SU(2)$-invariant Einstein manifolds.}
In the complete local classification of self-dual  $SU(2)$-invariant Einstein manifolds (see \cite{Hitc94}) 
the most non-trivial class
is described by the following metric on the unit ball:
\be
g=F\left\{\f{dx^2}{x(1-x)} +\f{\sigma_1^2}{\Omega_1^2} +\f{(1-x)\sigma_2^2}{\Omega_2^2}+
\f{x\sigma_3^2}{\Omega_3^2}\right\}\;.
\la{g1}\ee 
where $\sigma_j$ are left-invariant 1-forms on $SU(2)$ satisfying relations 
$d\sigma_j=\sigma_k\sigma_l$ for any even transposition $j,k,l$ of indeces $1,2,3$;
functions $F$ and $\Omega_j$ depend only on ``euclidean time'' $x$.

Self-duality of related Weyl tensor, together with assumption that this metric satisfies  Einstein equations
with cosmological constant $\Lambda$, implies (see \cite{Tod94}) system (\ref{eqOm}),  (\ref{int})
and the following formula for conformal factor $F$:
\be
F=-\f{1}{4 \Lambda}\f{8x\Omega_1^2\Omega_2^2\Omega_3^2 + 2\Omega_1\Omega_2\Omega_3
(x(\Omega_1^2+\Omega_2^2)-(1-4\Omega_3^2)
(\Omega_2^2-(1-x)\Omega_1^2))}{(x\Omega_1\Omega_2+2\Omega_3(\Omega_2^2-(1-x)\Omega_1^2))^2}\;.
\la{F}\ee

\item
{\it Three-dimensional Frobenius manifolds.}
Consider the following metric tensor on three-dimensional manifold $M$:
\be
g_0= -\Omega_1^2 d\g_1^2 +\Omega_2^2 d\g_2^2 +\Omega_3^2 d\g_3^2 
\la{Frob}\ee
Let us assume that the metric is flat i.e. associated Riemann tensor vanishes. Assume in addition that 
all variables $\Omega_j$ depend on $\g_j$ only via variable $x$ (\ref{x}). Then functions $\Omega_j(x)$
satisfy the system (\ref{eqOm}) (see \cite{Dubr92,Hitc97}). According to Dubrovin \cite{Dubr92}, 
in this case $M$ carries the structure of Frobenius manifold with spectrum $(-\f{1}{4},0,\f{1}{4})$ 
and metric (\ref{Frob}) 
corresponds to some solution of Witten-Dijkgraaf-Verlinde-Verlinde equation.
Variables $\g_j$ are called the canonical coordinates on $M$.

\end{enumerate}

It is worth to notice that all solutions of the system (\ref{eqOm}),  (\ref{int}) are known for at least several
years:
for example, they can be obtained via Okamoto transformations \cite{Okam87} from classical Picard solution of
Painlev\'e 6 equation; independently they were discovered in \cite{Hitc94}. However, the construction
of sect.2 allows to find the general solution of  (\ref{eqOm}),  (\ref{int}) in the following pretty
 simple form \cite{BabKor98}:
$$
\Omega_1= \f{e^{-\pi i p}}{2\pi i\th_2\th_4}\f{\f{d}{d q}\th\left[^{p+1/2}_{q+1/2}\right]}
{\th\left[^p_q\right]}\;,\hskip1.0cm
\Omega_2= \f{1}{2\pi\th_2\th_3}\f{\f{d}{d q}\th\left[^{p+1/2}_{\;\;\;q}\right]}{\th\left[^p_q\right]}\;,
$$
\be
\Omega_3=\f{e^{-\pi i p}}{2\pi i\th_3\th_4}\f{\f{d}{d q}\th\left[^{\;\;\;p}_{q+1/2}\right]}{\th\left[^p_q\right]}\;,
\la{solO}\ee
where 
\be
\th_2 \equiv \th\left[^{1/2}_{\;0}\right](0|\B)\;, \hskip0.5cm
\th_3 \equiv \th\left[^0_0\right](0|\B)\;, \hskip0.5cm
\th_4 \equiv \th\left[^{\;0}_{1/2}\right](0|\B)\;
\la{Jacobi}\ee
are standard theta-constants and $\th\left[^p_q\right]$ stands for $\th\left[^p_q\right](0|\B)$. 
Obviously, in the elliptic case period $\B$ depends only on the module $x$ of curve $\L$.

In addition to this general two-parametric family, there exist a one-parametric family of solutions
of (\ref{eqOm}),  (\ref{int}) which may be obtained from (\ref{solO}) in the limit $p,q\to 1/2$.

Substitution of expressions (\ref{solO}) into the formula for conformal factor (\ref{F}) gives the following 
simple result \cite{BabKor98}:
$$
F= -\f{2\pi^2\th_3^4}{\Lambda\f{d}{dq} {\rm log}\th\left[^p_q\right]}\Omega_1\Omega_2\Omega_3\;.
$$

In the paper \cite{BabKor98} formulas (\ref{solO}) were derived as a result of rather complicated calculation
starting from expression  (\ref{tau}) for the tau-function of Schlesinger system. However, they can also
 be derived in straightforward way 
from the formulas (\ref{Phi}), (\ref{phi1}), (\ref{psi1}), (\ref{Psi}) for function $\Psi(\g)$,
if in the process of  calculation of $\tr A_jA_k$ we choose free parameters $\g_\phi$ and $\g_\psi$ as 
follows: $\g_\phi=\g_j$, $\g_\psi=\g_k$.

\newpage

\end{document}